\documentstyle[11pt]{article}

\begin{document}
\title{\bf\Large Centaurus A as a source of extragalactic cosmic rays with
arrival energies well beyond the GZK cutoff}
\author{ Gustavo E. Romero $^{1,\, 2}$, Jorge A. Combi $^{1}$,\\
Santiago E. Perez 
Bergliaffa $^2$, Luis A. Anchordoqui $^2$}
\date{}
\maketitle
\noindent $^1$ Instituto Argentino de Radioastronom\'{\i}a, C.C. 5,
1894   Villa Elisa, Argentina\\ 
$^{2}$ Laboratorio de F\'\i sica Te\'orica, 
Departamento de F\'\i sica, U.N.L.P., 
C.C. 67,\\ 1900 La Plata, Argentina\\
\begin{abstract}

The ultra--high energy cosmic rays recently detected by several
air shower experiments could have an extragalactic origin. 
In this case, the nearest 
active galaxy Centaurus A might be the source of the most energetic
particles ever detected on Earth. We have used recent radio observations
in order to estimate the arrival energy of the protons accelerated by
strong shock fronts in the outer parts of this southern radio
source. We expect detections corresponding to particles with energies up to
$\sim 2.2 \times 10^{21}$ eV and an arrival direction  of ($l \approx
310^{\circ}$, $b \approx 20^{\circ}$) in galactic coordinates. The
future Southern Hemisphere Pierre Auger Observatory might provide a
decisive test for extragalactic models of the origin of the
ultra--high energy cosmic rays.
\end{abstract}

\newpage
\section{Introduction}

The origin of cosmic rays (CR) detected in the neighborhood of Earth
is still an unsettled issue. Most of the low energy particles are
believed to be produced by supernova explosions in the galactic 
interstellar medium \cite{1}. Higher energy CR could be accelerated 
by stellar explosions in strong stellar winds \cite{2}. 
The basic acceleration mechanism in all these models is diffuse 
shock acceleration \cite{3}.
Different observational facts, such as a break in the CR spectrum at
$\sim 5 \times 10^{18}$ eV, the change in the composition of the rays
from heavy nuclei below this break to light nuclei and protons at
higher energies, and the arrival directions of the highest energetic
particles seem to support the idea that the component with energies
below the break is predominantly galactic whilst the particles with
energies above $10^{19}$ eV have an extragalactic origin.
The hot spots of extended radio galaxies (i.e. the end--points  of
powerful jets ejected by the central active galactic nucleus) have
been suggested as the dominant source of ultra--high energy CR
\cite{5}. In such a case, particles are assumed to be accelerated 
by first order
Fermi mechanism through the strong, nonrelativistic shock waves
generated by huge plasma collisions.

When the ultra-high energy protons are injected in the intergalactic
medium, traveling losses originated by interactions with the 2.73 K
cosmic background radiation are expected to produce a high energy
cutoff in the particle spectrum at energies $\sim 10^{20}$ eV,
according to the predictions of Greisen \cite{6} and Zatsepin and
Ku\'zmin \cite{7} (the so--called ``GZK cutoff''). Nevertheless, 
recent detections with the Akeno Giant Air Shower Array
\cite{8} and the Utah Fly's Eye detector \cite{9,10} have revealed
the existence of CR with energies above the mentioned cutoff. A
detailed analysis of the arrival directions of these and other
energetic events has shown a correlation with the general direction
of the supergalactic plane, where many potential extragalactic
sources of CR are located \cite{11}. Despite these correlations, there
seems to be no conclusive evidence for an extragalactic origin of the most
energetic CR.  

In this paper we present a concrete candidate for testing the
acceleration models that involve hot spots and strong shock waves as
the source of the ultra--high energy component of the CR. Using
polarization measurements and synchrotron emission as a trace, 
we can make
an estimate of the maximum energy of the protons injected in the
extragalactic medium and the corresponding arrival energy for the case of the
nearest active galaxy: the southern radio source 
Centaurus A (Cen A)\footnote{Cen A was suggested as a possible source of
ultra-high energy cosmic rays by G. Cavallo \cite{cav}, from quite general 
energetic arguments.}. If the basic model is correct, 
a southern hemisphere
detector should register events produced by particles with energies
beyond $10^{21}$ eV as we shall see. But first, let us briefly
discuss some physical features of Cen A.

\section{Cen A and CR acceleration}

Cen A is a complex and extremely powerful radio source identified at
optical frequencies with the galaxy NGC 5128. It is the nearest radio
galaxy, at a distance of $\sim$ 3.5 Mpc according to recent works 
on globular clusters \cite{12} and planetary nebulae \cite{13}. Radio
observations at different wavelengths \cite{14,15} show a structure
composed by a compact core, a one--sided jet, Double Inner Lobes, a
Northern Middle Lobe, and two Giant Outer Lobes. This morphology,
together with the polarization data obtained by Junkes {\em et al.}
\cite{15} and the large--scale radio spectral index distribution
computed by two of us \cite{16}, strongly support the picture
of an active radio galaxy with a jet forming a relatively small angle
with the line of sight. The jet would be responsible for the
formation of the Northern Inner and Middle Lobes when interacting with
the interstellar and intergalactic medium, respectively. The Northern
Middle Lobe can be interpreted as a ``working surface'' \cite{17} at
the end of the jet, a place where strong shocks are produced by
plasma collisions, i.e. it can be considered as the hot spot of a
galaxy with a peculiar orientation.

The acceleration of particles in the hot spot is the result of repeated
scattering back and forth across a strong shock front in a partially
turbulent magnetic field. This process has been studied in detail by
Biermann and Strittmatter \cite{18}. Assuming that the energy density
per unit of wave number of MHD turbulence is of Kolmogorov type, i.e. 
$I(k) \; \alpha \; k^{-s}$ with $s= 5/3$, we have an acceleration time
scale for protons given by:
\begin{equation}
E_p \left( \frac{dE_p}{dt} \right)_{_{\rm ACC}}^{-1} \; \approx \; 
\frac{40}{\pi
\, c} \;\; \beta_{_{\rm JET}}\,\!\!\!\!\!\!\!^{-2} \;\; u \;\; 
R_{_{\rm HS}}\,\!\!\!\!\!^{2/3} \, \left(
\frac{E_p}{e \, B} \right)
\end{equation}
where $\beta_{_{\rm JET}}$ is the jet velocity in units of $c$, $u$ is
the ratio of turbulent to ambient magnetic energy density in the hot
spot (of radius $R_{_{\rm HS}}$), and $B$ is the total magnetic field
strength. The acceleration process will be efficient as long as the
energy losses by synchrotron radiation and photon--proton
interactions do not become dominant. Considering an average cross
section $\bar{\sigma}_{\gamma p}$ for the three dominant
pion--producing interactions \cite{19}: 
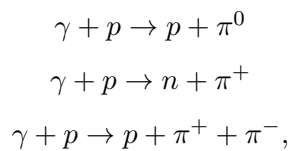
\begin{center}
$\gamma + p \rightarrow p + \pi^0$ \\
$\gamma + p \rightarrow n + \pi^+$ \\
$\gamma + p\rightarrow p + \pi^+ + \pi^- $,
\end{center}
we get the time scale of
the energy losses within a certainty of 80\%:
\begin{equation}
E_p  \left( \frac{dE_p}{dt} \right)_{_{\rm LOSS}}^{-1} \; \approx \;
\frac{6 \, \pi \, m_p^4 \, c^3}{\sigma_{_{\rm T}} \, m_e^2 \, B^2 \, 
(\,1 \,+\,
a\, A\,)}\; E_p^{-1}
\end{equation} 
where $a$ stands for the ratio of photon to magnetic energy densities, 
$ \sigma_{_T}$ is the classical Thomson cross section, 
and $A$ gives a measure of the relative strength of $\gamma p $
interactions  against the synchrotron emission. Biermann and
Strittmatter \cite{18} have estimated $A \approx 200$, almost
independently of the source parameters. The most energetic protons
injected in the intergalactic medium will have an energy that can be
obtained by balancing the energy gains and losses:
\begin{equation}
E_{p, \,{\rm max}} = 1259.3 \;\, c \;\, e^{1/4} \; \beta_{_{\rm JET}}\;
\!\!\!\!\!\!\!^{3/2}
\; \left( \frac{u}{\sigma_{_{\rm T}}} \right)^{3/4} 
\frac{m_p^2}{m_e^{3/2}} \;\, R_{_{\rm HS}}\,\!\!\!\!\!^{-1/2} \; 
B^{-5/4} \; ( 1 + A\,a)^{-3/4} \;\,{\rm MeV} \label{E}
\end{equation}
The acceleration model also predicts a power-law spectrum for the number of
particles per unit of energy: $N(E)\propto E^{-p}$, ($E<E_{p,\, {\rm max}}$).

In the case of Cen A, we can estimate $u$ from the radio spectral
index of the synchrotron emission in the Northern Middle Lobe and the
observed degree of linear polarization in the same region. We get $u
\approx 0.4$. The size of the hot spot can be directly measured
from the large--scale map obtained by Junkes {\em et al.} \cite{15}
with the assumed distance of 3.5 Mpc, giving as a result $R_{_{\rm HS}}
\approx  1.75$ kpc. 

The equipartition magnetic field can be obtained for the pre-shock region
from the detailed radio observations by Burns {\em et al} \cite{14}. The field 
component parallel to the shock will be amplified in the post-shock region by
a compression factor $\xi$. In the case of a strong, nonrelativistic shock 
front, $\xi\rightarrow 4$ \cite{lev}, and then, if 
$B_{\parallel}\sim B_{\perp}$, we have:
\begin{eqnarray}
B & \approx & (\xi ^{1/2} + 1 )^{1/2}\;B_{\perp} \nonumber \\
       & \approx & 5\times 10^{-5}\;\;\;{\rm G} \nonumber
\end{eqnarray}
An enhancement of the $B_{\parallel}$ component in the Northern Middle Lobe
can be clearly seen in the polarization maps displayed in ref. \cite{15}.
The value of $\beta_{_{\rm JET}}$ is uncertain. We shall assume 
$\beta_{_{\rm JET}}\approx 0.3$, a typical value for similar sources
\cite{20}. 

With the above mentioned values for the input parameters in Eq. (\ref{E}), 
the maximum energy of the protons injected in the intergalactic space
results 
\begin{equation}
E_{p,\,{\rm max}} \approx 2.7 \times 10^{21} \,\; {\rm eV}
\end{equation}
We can infer the index in the power-law spectrum from multifrequency 
observations of the synchroton radiation produced by the leptonic component of 
the particles accelerated in the source (see the standard
formulae, for instance, in the book by Pacholczyk \cite{pach}). 
Using the radio spectral index obtained by Combi and Romero
for the hot spot region \cite{16}, we get $p=2.2$.

\section{Propagation effects}

In this section we shall briefly discuss 
the proton energy losses due to collisions with microwave photons
during their travel to Earth, and the modifications in the energy 
spectrum arised in these interactions. The losses can be estimated as: 
\begin{equation}
\beta (E_p) = -\frac{1}{E_p}\frac{dE_p}{d\tau} = \frac{-kT}{2\pi ^2\hbar ^3 
c^2
\gamma_{p}^2} \sum_j\int_{\epsilon_{j,{\rm th}}}^\infty \epsilon
f_j (\epsilon) \ln\left [ 1 - \exp \left ( \frac{-\epsilon}{2\gamma _p kT}
\right ) \right ]\; d\epsilon
\label{pe}
\end{equation}
where $T$=2.73 K, $\tau$ is the traveling time, $f_j(\epsilon)$ is the cross 
section weighted by inelasticity of the $j$th reaction channel,
and the sum is carried out over all channels. The integration is
over the photon energy in the proton rest frame, from the threshold. For
energies higher than $3\times 10^{20}$ eV, and taking into account just the 
dominant pion-producing interactions, this equation becomes \cite{21}:
\begin{equation}
- \, \frac{1}{E_p} \, \frac{dE_p}{d\tau}\, \approx \, 1.8 \times 10^{-8}
\;\; {\rm yr}^{-1}
\end{equation}
According to this, the losses are
almost neglectable due to the proximity of the source, yielding a maximum 
arrival energy $E_{p,\, {\rm max}}^{\rm arr}\approx 2.2\times 10^{21}$ eV .
The arrival direction of these CR will be approximately ($l
\approx 310^{\circ}$, $b \approx 20^{\circ}$), in galactic coordinates.
Intergalactic and galactic magnetic fields with strength of the 
order of 
nG and $\mu$G, respectively, cause no relevant deviations from the line
of sight, due to the distances and energies involved in the case under 
consideration. However, a spreading in the arrival direction pattern should be
observed for a relatively large number of low energy events coming from
Cen A.

Both the $\gamma\pi$ energy losses and the conservation of the total number
of protons in the spectrum are expected to produce only a slight modification
of the original energy spectrum. This modification can be quantitatively 
described in terms of a factor given by
\begin{equation}
\eta (E_p,\tau) = \frac{N(E_p,\tau)}{N(E_p,0)}
\end{equation}
In the case of a nearby source, an analitical expression of $\eta (E_p,\tau)$ 
has been found by Berezinsky and Grigor'eva \cite{21}:
\begin{equation}
\eta(E_p,\tau) = 1 + \beta(E_p)\, \tau \left ( \frac{d\ln \beta(E_p)}{d\ln E_p} 
- p + 1 \right )
\end{equation}
At very high energies ($E_p >5\times 10 ^{20}$), $d\ln\beta(E_p)/d\ln E_p
\rightarrow 0$, and then $\eta(E_p,\tau) \rightarrow 1-(p-1)\tau\omega _0$,
where $\omega_0\approx 1.8\times 10^{-8}$ yr$^{-1}$. Since $\eta<1$, there 
will be a slight fall in the spectrum, as we already mentioned in the case of 
the highest energy protons.
At lower energies ($E_p\stackrel{<}{\sim}3\times 10^{20}$ eV) the modification
factor is
\begin{equation}
\eta(E_p,\tau) \approx 1 + \left ( \frac{\tau}{\tau_\pi} \right )
e^{-(\epsilon _\pi /E_p) }\left ( \frac{\epsilon _\pi}{E_p} - p + 1 \right)
\end{equation}
where $\epsilon_\pi\approx 3\times 10^{20}$ eV, and $\tau _\pi \approx 
1.7\tau$. 
This will produce a small bump in the spectrum. In brief, the overall
spectrum will be scarcely modified.

An additional interesting consequence of the model is that 
the \mbox{$\gamma + p \rightarrow p +
\pi^0$} interactions would be a source  of $\gamma$--rays
through $\pi^0$ decay \mbox{($\pi^0 \rightarrow 2 \, \gamma$)}. 
Considering a power-law dependence for the isotropic and homogeneous photon
distribution in the source \cite{18}:
\begin{equation}
N_{\gamma}(\epsilon)= n_{0}\left(\epsilon\over \epsilon_{0}\right)^{-2}
\;\;\;\;\;\;\;\epsilon_0\leq\epsilon\leq\epsilon_{\rm max}
\end{equation}
we can compute the energy range of the total pion spectrum \cite{man}. Using 
$\gamma_{p,{\rm max}}\approx10^{12}$ and $\epsilon_{\rm max}\approx 1$ MeV,
we obtain (in the observer's frame):
\begin{equation}
E_{\pi}\in [4\times 10^{10} {\rm eV},\; 5\times 10^{20} {\rm eV}]
\label{inter}
\end{equation}
If the proton distribution $N(E_p)$ has the same index as the electron
distribution $N(E_e)$, as we have assumed, the pion distribution function will 
be
\begin{equation}
N(E_\pi)\propto E_{\pi}^{-\frac{p+1}{2}}
\end{equation}
In the case of Cen A, we have $N(E_\pi)\propto (\gamma_\pi\;m_\pi
\;c^2)^{-1.6}$.  

Emission at low $\gamma$-ray frequencies from Cen A has been detected by the 
Compton Gamma
Ray Observatory \cite{22}, and by experiments with balloons \cite{23}. However,
these $\gamma$-rays could not be produced by $\pi^0$ decays in the acceleration
region because, according to (11), the relevant energy ranges from $\sim 20$ 
GeV up to a value comparable to those of ultra-high energy CR. 
It might be worthwhile then to perform observations of Cen A with
instruments sensitive to TeV energies, for instance with the new
Imaging Atmospheric \u{C}erenkov Telescopes \cite{aha}.

\section{Final comment}

We expect the detection of an extensive air shower produced 
by the collision of protons with energies \mbox{$E_p \,>\, 100$} EeV with 
the Earth's atmosphere. 
Since the arrival direction of these extremely energetic particles is well
known, the future Hybrid Pierre Auger Observatory \cite{24,25} will provide a
decisive test for the models that invoke an extragalactic origin for
the ultra-high energy component of the cosmic rays.   

\vspace{1.0cm}

\noindent Remarks by Alan Watson are gratefully acknowledged.
This work has been partially supported by CONICET and UNLP.


\begin{thebibliography}{99}
\bibitem{1} R. D. Blandford and J. P. Ostriker, {\em Ap.J.} {\bf
 237}, 793 (1980).
\bibitem{2} H. J. V\"{o}lk and P. L. Biermann, {\em Ap. J. Letters} {\bf
 333}, L65 (1988).
\bibitem{3} L. O'C. Drury, {\em Rep. Prog. Phys.} {\bf 46}, 973 (1983).
\bibitem{5} J. P. Rachen and P. L. Biermann, {\em Astron. Astrophys.}
 {\bf 272}, 161 (1993).
\bibitem{6} K. Greisen, {\em Phys. Rev. Lett.} {\bf 16}, 748 (1966).
\bibitem{7} G. T. Zatsepin and V. A. Ku\'zmin, {\em Pis'ma Zh. Eksp.
 Teor. Fiz.} {\bf 4}, 114 (1966).
\bibitem{8} N. Hayashida, {\em et al.}, {\em Phys. Rev. Lett.} {\bf
 73}, 3491 (1994).
\bibitem{9} D. J. Bird, {\em et al.}, {\em Phys. Rev. Lett.} {\bf
 71}, 3401 (1993).
\bibitem{10} D. J. Bird, {\em et al.}, {\em Ap. J.} {\bf 441}, 144 (1995).
\bibitem{11} T. Stanev, P. L. Biermann, J. Lloyd-Evans, J. P. Rachen 
 and A. A. Watson, {\em Phys. Rev. Lett.}, {\bf 73}, 3056 (1995).
\bibitem{cav} G. Cavallo, {\em Astron. Astrophys.} {\bf 269}, 45 (1978).  
\bibitem{12} G. L. H. Harris, {\em et al.}, {\em Ap. J.} {\bf 287},
 175 (1984).
\bibitem{13} G. H. Jacobi, R. Ciardullo and H. Ciford, in {\em The
 Extragalactic Distance Scale}, ASP Conf. Ser. 4, San Francisco, p.42 (1988).
\bibitem{14} J. O. Burns, E. D. Feigelson and E. J. Schreier, {\em
 Ap. J.} {\bf 273}, 128 (1983).
\bibitem{15} N. Junkes, {\em et al.}, {\em Astron. Astrophys.} {\bf
 269}, 29 (1993).
\bibitem{16} J. A. Combi and G. E. Romero, {\em Astron. Astrophys.},
 submitted.
\bibitem{17} R. D. Blandford and M. J. Rees, {\em Monthly Not. Royal
 Astron. Soc.} {\bf 169}, 395 (1974).
\bibitem{18} P. L. Biermann and P. A. Strittmatter, {\em Ap. J.} {\bf
 322}, 643 (1987).
\bibitem{19} T. A. Armstrong, {\em et al.}, {\em Phys. Rev.} D {\bf
 5}, 1640 (1972).
\bibitem{lev} L. Landau and E. Lifchitz, {\em Fluid Mechanics} (Pergamon Press,
 Oxford, 1958).
\bibitem{20} K. Meisenheimer, {\em et al.}, {\em Astron. Astrophys.}
 {\bf 219}, 63 (1989).
\bibitem{pach} A. G. Pacholczyk, {\em Radio Astrophysics} (Freeman, San
 Francisco, 1970).
\bibitem{21} V. S. Berezinsky and S. I. Grigor'eva, {\em Astron.
 Astrophys.} {\bf 199}, 1 (1988).
\bibitem{man} K. Mannheim and P. L. Biermann, {\em Astron. Astrophys.} 
 {\bf 221}, 211 (1989).
\bibitem{22} R. L. Kinzer, {\em et al.}, {\em Ap. J.} {\bf 449}, 105 (1995).
\bibitem{23} P. von Ballmoos, R. Diehl and V. Sch\"{u}nfelder, {\em
 Ap. J.} {\bf 312}, 134 (1987). 
\bibitem{aha} F.A. Aharonian, {\em Nucl. Phys. B (Proc. Suppl.)} {\bf 39A}, 
 193 (1995).
\bibitem{24} J. W. Cronin, {\em The Highest Energy Particles Produced in 
 the Universe: Cosmic Rays}, talk delivered at the Auger Project Workshop, 
 Bariloche, Argentina, October 1995. 
\bibitem{25} A. A. Watson, {\em The Highest Energy Cosmic Rays: Recent
 Measurements and their Instrumentation}, talk delivered at the Auger
 Project Workshop, Bariloche, Argentina, October 1995.
\end{thebibliography}
\end{document}